# Consideration on new functionality based on photo-excitation of magnetization in ultra-short time region

H. Munekata

Tokyo Institute of technology, 4259-J3-15 Nagatsuta, Midori-ku, Yokohama 226-8503, Japan

e-mail address: hiro@isl.titech.ac.jp

**Abstract** New functionalities on the basis of photo-induced, non-equilibrium magnetism are discussed in the context of optical integrated circuit. Importance of studies in weak excitation regime is stated, referring experimental results obtained by ultrafast spectroscopy using Co/Pd mutilayers.

## Introduction

Optical integrated circuit on a chip has become a matter of serious target in research and development. On the chip, dielectric waveguides bear optical interconnects among semiconductor-based active devices. The turn of magnetic materials will be in a waveguide-type optical isolator, which is natural and reasonable extension in view of achievement realized by bulk-type optical isolators. In this paper, I wish to discuss new functionalities that would appear when externally-controllable magnetism is added on the waveguides.

## New functionalities based on photo-excitation of magnetization

Approach to this interesting thought is open at present. I wish to mention two efforts taking place in my laboratory: firstly, magneto-optical (MO) polarization modulation of light in a waveguide, and secondly, magneto-optical phase delay of light in a waveguide. The first subject of a magneto-optical modulator aims at multiplexed transmission of polarization modulation signals which transmit in a multi-mode waveguide. The most intriguing idea underlying this device is that local modulation of magnetic state can give rise to modulation of polarization of selected modes of light in a waveguide. See ref.1 which reported the demonstration of this idea using a magnet-fiber hybrid structure. The idea of the second work is based on a MO isolator utilizing Mach-Zehnder interferometer (MZI) [2,3], in which the light coupled into an input port X is guided into two branches, undergoes MO phase delay $\Delta\phi$ for the light in one branch, interferes at the yielding point of the two branch Y, and comes out from an output port Z, as shown schematically in Fig.1. The light energy at the port Z can be tuned between 0 (destructive interference, $\Delta\phi = \pi$) and 1 (constructive interference, $\Delta\phi = 0$) depending on the value of $\Delta\phi$; namely, the direction of a magnetization vector *M* of a magnetic layer M. The MZI-MO isolator relies on a large imbalance in light energy between the light propagating from X to Z and vice versa with a fixed *M* due to the non-reciprocity. Suppose now that *M* can be tuned externally. In such a case, intensity of light from X to Z can be modulated as a function *M(t)*. This gives rise to various new functionalities, such as optical version of MRAM if one utilizes non-volatility of magnetization, and optical version of a transistor if one utilizes ultrafast non-equilibrium magnetic states caused by optical excitations.

Let us suppose an intense a seed light pulse train R of pulse width and interval, respectively, $\Gamma$ and $1/\Omega$, entering a port X, and a relatively weak signal-carrier pulse train S of the same $\Gamma$ and $1/\Omega$ (8 bit, 10011001) impinging a magnetic layer M. When *M* responds synchronously with S, the interference condition at Y is also modulated accordingly, thus yielding an intense, pulse train R* (10011001) from an outport Z. With this protocol, a sort of transistor function can be achieved. In general, the data transmission rate *r* can be enhanced significantly by reducing $\Gamma$ within the context of the digital baseband transmission [4]. This notion addresses new, long-term challenges concerning developments of fs-lasers and highly sensitive detectors on a chip.

There are a few requirements in view of properly and efficiently copying information carrier signals S on M. Firstly, initial response of magnetization should be fast enough, within the pulse width $\Gamma$; secondly, recovery of magnetization should be faster than the pulse interval $1/\Omega$; and thirdly, energy per pulse of S should be less than that of R. With those criteria in mind, photo-excited precession of magnetization (PEPM) in the regime of weak excitation (< 10 $\mu J/cm^2$/pulse) has been studied using (Ga,Mn)As and ultrathin Co/Pd multilayers (MLs). In the next section, recent progress in the study on PEPM with Co/Pd MLs is reviewed, through which a mechanism that enhances efficiency of photo-excitation is concisely addressed.

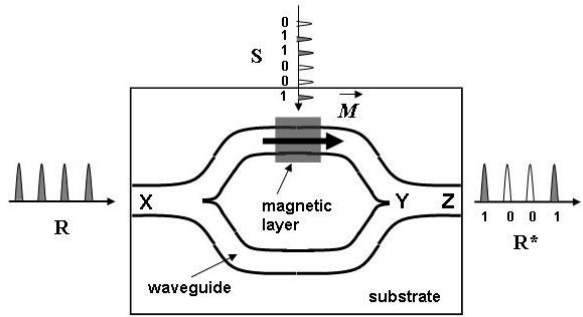

Fig. 1. A schematic illustration of an active three-terminal, magnet-waveguide hybrid structure based on MZI.

## 2. PEPM in Co/Pd MLs around 1 ps time region

It was found that PEPM in Co/Pd MLs could be triggered by fs-laser pulses of the fluence as low as around 1 μJ/cm$^2$ [5]. This fact implies that the mechanism other than ultrafast demagnetization may have significant influence on PEPM. Shown in Fig. 2 are temporal profiles of PEPM in the [Co(0.78 nm) / Pd(0.81 nm)]$_5$ sample with various pump fluences $F$. Amplitude of oscillation increases throughout the entire time region with increasing $F$. Sharp spikes, being attributed to the ultrafast demagnetization, also develops at $t < 10$ ps for relatively high $F$. Profiles in the region $t > 20$ ps are well fitted with the LLG equation, through which we find a temporal profile of the effective field, $H(t) = \{1 + D \exp(-t/\tau)\} \cdot |H_{ani} - H_{dem}|$, where $H_{ani}$ and $H_{dem}$ are perpendicular anisotropy field and demagnetization field, respectively, and $D$ the magnitude of a sudden change upon pulsed excitation and $\tau$ its lifetime. $D$ varies linearly with $F$, whereas $\tau$ is nearly constant $\tau \approx 400$ ps. Note that not only $H_{dem}$ but also $H_{ani}$ is influenced simultaneously by the pulsed laser excitation. Furthermore, discrepancy between experiment and LLG at $t < 20$ ps indicates that the tip-off angle of a magnetization vector at the re-magnetized state ($\tau \sim 5$ ps) is larger than that expected by precession dynamics, suggesting that the energy supplied by the photonic excitation is consumed in part for varying the tip-off angle of magnetization even in the process of re-magnetization [6].

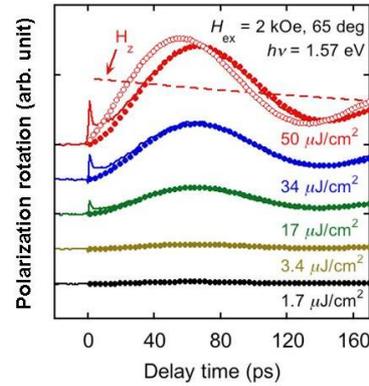

**Fig. 2.** Experimental PEPM data (solid lines) and calculated values (dots).

**Acknowledgments** We acknowledge partial supports from Advanced Photon Science Alliance Project from MEXT and Grant-in- Aid for Scientific Research (No. 22226002) from JSPS.